\documentstyle[12pt,epsf]{article}

\begin{document}

\baselineskip 6mm
\renewcommand{\thefootnote}{\fnsymbol{footnote}}


\newcommand{\nc}{\newcommand}
\newcommand{\rnc}{\renewcommand}


\rnc{\baselinestretch}{1.24}    
\setlength{\jot}{6pt}       
\rnc{\arraystretch}{1.24}   

\makeatletter
\rnc{\theequation}{\thesection.\arabic{equation}}
\@addtoreset{equation}{section}
\makeatother



\nc{\be}{\begin{equation}}

\nc{\ee}{\end{equation}}

\nc{\bea}{\begin{eqnarray}}

\nc{\eea}{\end{eqnarray}}

\nc{\xx}{\nonumber\\}

\nc{\ct}{\cite}

\nc{\la}{\label}

\nc{\eq}[1]{(\ref{#1})}

\nc{\newcaption}[1]{\centerline{\parbox{6in}{\caption{#1}}}}

\nc{\fig}[3]{

\begin{figure}
\centerline{\epsfxsize=#1\epsfbox{#2.eps}}
\newcaption{#3. \label{#2}}
\end{figure}
}


\def\CA{{\cal A}}
\def\CC{{\cal C}}
\def\CD{{\cal D}}
\def\CE{{\cal E}}
\def\CF{{\cal F}}
\def\CG{{\cal G}}
\def\CH{{\cal H}}
\def\CK{{\cal K}}
\def\CL{{\cal L}}
\def\CM{{\cal M}}
\def\CN{{\cal N}}
\def\CO{{\cal O}}
\def\CP{{\cal P}}
\def\CR{{\cal R}}
\def\CS{{\cal S}}
\def\CU{{\cal U}}
\def\CW{{\cal W}}
\def\CY{{\cal Y}}


\def\IR{{\hbox{{\rm I}\kern-.2em\hbox{\rm R}}}}
\def\IB{{\hbox{{\rm I}\kern-.2em\hbox{\rm B}}}}
\def\IN{{\hbox{{\rm I}\kern-.2em\hbox{\rm N}}}}
\def\IC{\,\,{\hbox{{\rm I}\kern-.59em\hbox{\bf C}}}}
\def\IZ{{\hbox{{\rm Z}\kern-.4em\hbox{\rm Z}}}}
\def\IP{{\hbox{{\rm I}\kern-.2em\hbox{\rm P}}}}
\def\IH{{\hbox{{\rm I}\kern-.4em\hbox{\rm H}}}}
\def\ID{{\hbox{{\rm I}\kern-.2em\hbox{\rm D}}}}


\def\a{\alpha}
\def\b{\beta}
\def\ga{\gamma}
\def\d{\delta}
\def\ep{\epsilon}
\def\ph{\phi}
\def\k{\kappa}
\def\l{\lambda}
\def\m{\mu}
\def\n{\nu}
\def\th{\theta}
\def\rh{\rho}
\def\s{\sigma}
\def\t{\tau}
\def\w{\omega}
\def\G{\Gamma}


\def\half{\frac{1}{2}}
\def\dint#1#2{\int\limits_{#1}^{#2}}
\def\goto{\rightarrow}
\def\para{\parallel}
\def\brac#1{\langle #1 \rangle}
\def\grad{\nabla}
\def\curl{\nabla\times}
\def\div{\nabla\cdot}
\def\p{\partial}
\def\e{\epsilon_0}


\def\Tr{{\rm Tr}\,}
\def\det{{\rm det}}


\def\vare{\varepsilon}
\def\barz{\bar{z}}
\def\barw{\bar{w}}


\def\ad{\dot{a}}
\def\bd{\dot{b}}
\def\cd{\dot{c}}
\def\dd{\dot{d}}
\def\so{SO(4)}
\def\sop{SO(4)^\prime}
\def\bc{{\bf C}}
\def\bfz{{\bf Z}}
\def\bz{\bar{z}}

\begin{titlepage}

\hfill\parbox{5cm} {SOGANG-HEP 304/03 \\
AEI-2003-018 \\
{\tt hep-th/0302060}}

\vspace{25mm}

\begin{center}
{\Large \bf Superstrings and D-branes in A Plane Wave}

\vspace{15mm}
Jongwook Kim$^{\, a \,}$\footnote{jongwook@sogang.ac.kr},
Bum-Hoon Lee$^{\, a,b \,}$\footnote{bhl@ccs.sogang.ac.kr}
and Hyun Seok Yang$^{\, a \,}$\footnote{hsyang@hepth.sogang.ac.kr}
\\[10mm]

${}^a$ {\sl Department of Physics, Sogang University,
Seoul 121-742, Korea} \\
${}^b$ {\sl Max-Planck-Institut f\"ur Gravitationsphysik, \\
Albert-Einstein-Institut, Am M\"uhlenberg 1, D-14476 Golm,
Germany}
\end{center}

\thispagestyle{empty}

\vskip2cm


\centerline{\bf ABSTRACT}
\vskip 4mm
\noindent

We carefully analyze the supersymmetry algebra of closed strings
and open strings in a type IIB plane wave background. We use eight
component chiral spinors, $SO(8)$ Majorana-Weyl spinors, in
light-cone gauge to provide a useful basis of string field theory
calculation in the plane wave. We consider the two classes of
D-branes, $D_\pm$-branes, and give a worldsheet derivation of conserved
supercurrents for all half BPS D-branes preserving 16
supersymmetries in the type IIB plane wave background. We
exhaustively provide the supersymmetry algebra of the half BPS
branes as well. We also point out that the supersymmetry
algebra distinguishes the two $\so$ directions with relative sign
which is consistent with the $Z_2$ symmetry of the string action.
\\


\vspace{1cm}

\today

\end{titlepage}

\renewcommand{\thefootnote}{\arabic{footnote}}
\setcounter{footnote}{0}

\section{Introduction}

The Penrose limit of the $AdS_5 \times S^5$ background in type
IIB supergravity corresponds to a plane wave solution \ct{blau},
\bea \la{pp-metric}
&& ds^2 = -2 dx^+ dx^- - \mu^2 x_I^2 (dx^+)^2 + {dx_I}^2, \\
&& F_{+1234}=F_{+5678}=2 \mu. \nonumber
\eea
This implies a correspondence between type IIB string theory in
the plane wave background \eq{pp-metric} and $\CN=4$
supersymmetric Yang-Mills theory with large R-charge, in a sense
as a part of $AdS_5/CFT_4$ correspondence. Since the background
\eq{pp-metric} is one of the very few Ramond-Ramond backgrounds on
which string theory is exactly solvable \ct{metsaev1,metsaev2},
one may have a genuine
hope to explicitly check the conjectured AdS/CFT correspondence
beyond the supergravity approximation on the string theory side.
Indeed Berenstein, Maldacena and Nastase \ct{bmn} succeeded in
reproducing the string spectrum from perturbative super Yang-Mills
theory, thereby putting the duality on a firm ground at the free
theory level. Subsequent developments using the super Yang-Mills
theory \ct{kris}-\ct{gursoy} and the light-cone string
field theory \ct{spradlin1}-\ct{kiem2} showed
that the duality is still valid even after the
interactions both on the super Yang-Mills theory side and on the
string theory side are introduced.

D-branes can be described by boundary states of closed string
state. The symmetries that the boundary state preserves are thus
the combinations of the closed string symmetries that leave the
boundary state invariant. Recently possible D-branes in the plane
wave background \eq{pp-metric} were identified
and their supersymmetries were classified \ct{billo}-\ct{gaberdiel}.
In particular, Skenderis and Taylor showed in \ct{skenderis1,skenderis2}
that, although the kinematical supersymmetry descending from
the closed string is totally broken on a $D_+$-brane \ct{hikida,gaberdiel}
(see section 2 for the definition of $D_\pm$-branes),
a different kind of kinematical supersymmetry is nontrivially
realized by incorporating the worldsheet symmetries. We will study
this kinematical supersymmetry too.

Since the plane wave background \eq{pp-metric} allows a light-cone
gauge choice and the string theory in this background takes the
simplest form in the light-cone gauge, the most straightforward
method for constructing the superstring interactions is to use the
light-cone Green-Schwarz formalism \ct{metsaev1}.
Since the original papers \ct{green1,green2,green3}
on string field theory were using the 8-component spinors, $SO(8)$
Majorana-Weyl spinors, it may be desirable to provide the
supersymmetry algebra in this basis as a useful reference for
string field theory calculation in the plane wave, although
detailed analysis had been done in \ct{metsaev1,metsaev2} for the closed
string and in \ct{skenderis1,skenderis2} for the open string
using 16-component spinors, $SO(9,1)$ Majorana-Weyl
spinors. Thus we will exhaustively provide the supersymmetry
algebras for both closed strings and open strings in the plane
wave background using the 8-component spinors.

As is well known the spectrum of light-cone hamiltonian of plane
wave superstring is discrete. As analyzed in \ct{metsaev3}
recently, the super Yang-Mills theory in a four-dimensional plane
wave background has also a discrete spectrum of light-cone energy
operator, so expected to be most appropriate for establishing a
precise correspondence with plane wave superstring and for
studying the holographic issues of the duality. Since the plane
wave super Yang-Mills theory is the low energy worldvolume action
of probe D3-branes in the plane wave background \eq{pp-metric}
and is the ground-state sector of the open string field theory
which also has a discrete energy spectrum, it is plausible that
the plane wave super Yang-Mills theory including interactions can
be captured by the open superstring field theory defined on the
probe D3-branes \ct{kim}. Furthermore the open-closed
string duality \ct{bergman,gaberdiel},
being, probably, an underlying principle of
AdS/CFT duality, implies that the open string field theory
possibly reproduces the corresponding results of closed string
field theory. For this purpose it will be useful to explicitly
construct the supersymmetry algebra for the open
string in the plane wave background \eq{pp-metric}
using 8-component spinors mostly used in the light-cone string
field theory \ct{green1,green3}.

This paper is organized as follows. In Sec. 2, we discuss the
canonical quantization of superstrings in the plane wave
background and present the mode expansions of open strings
compatible with boundary conditions. In Sec. 3, we first analyze
the supersymmetry algebra of closed string to provide its explicit
expressions in terms of 8-component spinors. We find that the
supersymmetry algebra distinguishes the two $\so$ directions
with opposite sign consistent with the $Z_2$ symmetry of the
string action \ct{chu1}. We derive conserved
supercurrents for all half BPS D-branes preserving 16
supersymmetries in the type IIB plane wave background and
exhaustively provide the mode expansion of symmetry generators and
supersymmetry algebra for half BPS D-branes, including
D-branes with worldvolume flux which were not studied in \ct{skenderis1,skenderis2},
using 8-component spinors. In Sec. 4, we briefly review our
results and address some other issues. In Appendix A, we explain
our notations and definitions and give useful formula used in this
paper. In Appendix B, we provide some technical details on
(anti-)commutation relations for $D_+$-branes. In Appendix C, we
present how the kinematical supersymmetry of $D_+$-brane can be
derived from the open string mode expansion.

\section{Canonical Quantization of Superstrings in Plane Wave}

The Green-Schwarz light-cone action in the plane wave background \eq{pp-metric}
describes eight free massive bosons and fermions. In the
light-cone gauge, $X^+=\tau$, the action is given by
\be \la{gs-action}
S = \frac{1}{2\pi \a^\prime p^+} \int d\tau \int_{0}^{2\pi \a^\prime |p^+|}
d\sigma \Bigl[ \half \p_+X_I \p_-X_I - \half \mu^2 X_I^2
- i \bar{S}(\rho^A \p_A - \mu \Pi)S \Bigr]
\ee
where $\p_\pm = \p_\tau \pm \p_\sigma$.
Our notations and conventions are summarized in Appendix A.
In this paper we will take $\a =\a^\prime p^+$ for
closed string and $\a =2 \a^\prime p^+$ for open string. We take
the spinor $S$ as eight two-component Majorana spinors on the worldsheet $\Sigma$
that transform as positive chirality spinors ${\bf 8}_s$
under $SO(8)$\footnote{The spinors
$S^A$ are two $SO(8)$ Majorana-Weyl spinors satisfying the
light-cone gauge. These spinors can be obtained by the original
$SO(9,1)$ Majorana-Weyl spinors $\theta^A$ satisfying
$\bar{\gamma}^+\theta^A=0$,
fixing the $\kappa$-symmetry. As in \eq{so8-spinor},
the fermionic light-cone gauge can be solved by taking
$S^A=- \half \gamma^+ \bar{\gamma}^- \theta^A$.}:
\be \la{ws-s}
S^a= \left(%
\begin{array}{c}
  S^{1a} \\
  S^{2a} \\
\end{array}%
\right), \qquad
\bar{S}^a= {S^a}^T \rho^\tau,
\ee
where
\be \la{2d-dirac}
\rho^\tau=\left(%
\begin{array}{cc}
  0 & -1 \\
  1 & 0 \\
\end{array}%
\right), \qquad
\rho^\sigma =\left(%
\begin{array}{cc}
  0 & 1 \\
  1 & 0 \\
\end{array}%
\right).
\ee
The presence of $\Pi$ in the fermionic action breaks the symmetry
from $SO(8)$ to $\so \times \sop$.
The equations of motion following from the action \eq{gs-action}
take the form
\bea \la{eom-boson}
&& \p_+\p_-X^I + \mu^2 X^I =0, \\
\la{eom-fermion}
&& \p_+ S^1 - \mu \Pi S^2 =0, \qquad \p_-S^2 + \mu \Pi S^1 =0.
\eea

\subsection{Closed string}

First we analyze the closed string case. Detailed analysis
using 16-component spinors had been done in \ct{metsaev1,metsaev2}
for the closed string and in \ct{skenderis1,skenderis2} for open strings.
We will instead use 8-component spinors
as in the original papers \ct{green1,green2,green3}
on string field theory. This will be a useful basis for light-cone
string field theory and for open string analysis.

The general solutions to Eqs. \eq{eom-boson} and \eq{eom-fermion}
for the closed string are found to be\footnote{In our following analysis
the mode expansion of string fields will be performed for real
fields. Thus the reality of the field requires that $\xi_n^\dagger = \xi_{-n}$
for any bosonic or fermionic n-th mode $\xi_n$.}
\bea \la{sol-boson}
&& X^I(\sigma, \tau)=\cos\mu\tau x_0^I + \sin\mu\tau \frac{p_0^I}{\mu} +i\sum_{n
\neq 0}\frac{1}{\omega_n}(\varphi_n^1(\sigma,\tau) \a_n^{1I} + \varphi_n^2(\sigma,\tau)
\a_n^{2I}), \\
\la{sol-fermion}
&& S^1 (\sigma,\tau)= \cos \mu\tau S_0^1
+\sin\mu\tau \Pi S_0^2
+ \sum_{n \neq 0}c_n(\varphi_n^1(\sigma,\tau) S_n^1
+i \rho_n \varphi_n^2(\sigma,\tau)\Pi S_n^2), \xx
&& S^2 (\sigma,\tau)= \cos \mu\tau S_0^2
- \sin\mu\tau \Pi S_0^1
+ \sum_{n \neq 0}c_n(\varphi_n^2(\sigma,\tau) S_n^2
- i \rho_n \varphi_n^1(\sigma,\tau)\Pi S_n^1),
\eea
where the basis functions $\varphi_n^{1,2}(\sigma,\tau)$ are
defined by
\be \la{basis}
\varphi_n^1(\sigma,\tau)=e^{-i(\omega_n \tau -
\frac{n}{|\a|}\sigma)}, \qquad \varphi_n^2(\sigma,\tau)=e^{-i(\omega_n \tau
+ \frac{n}{|\a|}\sigma)}
\ee
and
\be \la{omega-etc}
\omega_n = \mbox{sgn}(n) \sqrt{\mu^2 + n^2/\a^2}, \quad \rho_n=
\frac{\omega_n-n/|\a|}{\mu}, \quad c_n= \frac{1}{\sqrt{1+\rho_n^2}}.
\ee

For canonical quantization let us introduce the canonical momenta
\be \la{momenta}
P^I(\sigma,\tau)= \frac{1}{2\pi|\a|}\dot{X}^I(\sigma,\tau),
\qquad P^{Aa}(\sigma,\tau)= \frac{i}{2\pi|\a|}S^{Aa}(\sigma,\tau),
\ee
where $\dot{X}^I=\frac{\p x^I}{\p \tau}$.
After removing the second class constraint in the fermionic part
by introducing the Dirac bracket, we get the following equal-time
quantum (anti-)commutation relations
\bea \la{commutator-field-boson}
&&[X^I(\sigma,\tau), \dot{X}^J(\sigma^\prime,\tau)]=
i2\pi|\a| \delta^{IJ} \delta(\sigma-\sigma^\prime),\\
\la{commutator-field-fermion}
&& \{S^{Aa}(\sigma,\tau), S^{Bb}(\sigma^\prime,\tau) \}=
\pi|\a| \delta^{ab} \delta^{AB} \delta(\sigma-\sigma^\prime).
\eea
The above quantization gives the (anti-)commutation relations for
the modes in \eq{sol-boson} and \eq{sol-fermion}:
\bea \la{commutator-mode-boson}
&& [x_0^I, p_0^J]=i\delta^{IJ}, \qquad [\a_n^{IA},\a_m^{JB}]= \half \omega_n
\delta_{m+n,0} \delta^{IJ}\delta^{AB}, \\
\la{commutator-mode-fermion}
&& \{S_n^{Aa}, S_m^{Bb} \} = \frac{1}{2} \delta_{n+m,0}
\delta^{ab}\delta^{AB}.
\eea

\subsection{Open string}

Now we will discuss open strings living on a D-brane in the plane
wave background \eq{pp-metric}. Here we consider only static
D-branes for simplicity. To describe a Dp-brane, we impose the
Neumann boundary conditions on $(p-1)$ coordinates and Dirichlet
boundary conditions on the remaining transverse coordinates:
\bea \la{bc-boson}
&& \p_\sigma X^r|_{\p\Sigma}=0, \\
&& \p_\tau X^{r^\prime}|_{\p\Sigma}=0.
\eea
For the fermionic coordinates, the appropriate boundary condition
\ct{lambert}
is
\be \la{bc-fermion}
S^1|_{\p\Sigma}=\Omega S^2|_{\p\Sigma}.
\ee

It turns out \ct{skenderis1,skenderis2,gaberdiel}
that in the plane wave background there are two
classes of maximally supersymmetric Dp-branes, depending on the
choice of $\Omega$:
\be \la{two-class}
D_-: \Pi\Omega \Pi\Omega = -1, \qquad D_+: \Pi\Omega\Pi\Omega =1.
\ee
The boundary conditions, $D_\pm$, in \eq{two-class} together with the
fermionic equations of motion, \eq{eom-fermion}, imply \ct{skenderis1} that
\bea \la{bc-sigma1}
&& D_-: \p_\sigma S^1|_{\p\Sigma} = - \Omega \p_\sigma S^2|_{\p\Sigma}, \\
\la{bc-sigma2}
&& D_+: \p_\sigma S^1|_{\p\Sigma} = (-\Omega \p_\sigma S^2 + 2 \mu \Pi
S^2)|_{\p\Sigma}.
\eea
The boundary condition for $X^-$ is determined by the Virasoro
constraint
\be \la{virasoro}
\p_\sigma X^- = \p_\tau X^I \p_\sigma X^I -\frac{i}{2}(S^1\p_+S^1-
S^2\p_-S^2),
\ee
from which one can see that the $X^-$ coordinate must satisfy the
Neumann boundary condition for both classes in \eq{two-class}.

Here we list the allowed choices for $\Omega$ consistent with each
constraint in Eq. \eq{two-class}. For $D_-$-branes, there are the following
possibilities \ct{dabholkar,sken-tayl, skenderis1,gaberdiel}
(for the notation used below, see Appendix A):
\bea \la{d-brane}
&& D3: (m,n)\; =\;(2,0),\; (0,2), \xx
&& D5: (m,n)\; =\;(3,1),\; (1,3), \\
&& D7: (m,n)\; =\;(4,2),\; (2,4). \nonumber
\eea
For $D_+$-branes, there are the following possibilities \ct{sken-tayl,skenderis1,gaberdiel}:
\bea \la{d+brane}
&& D1: (m,n)\; =\;(0,0), \xx
&& D3: (m,n)\; =\;(1,1), \xx
&& D5: (m,n)\; =\;(4,0),\; (2,2),\;(0,4), \\
&& D7: (m,n)\; =\;(3,3), \xx
&& D9: (m,n)\; =\;(4,4). \nonumber
\eea

The quantization of open strings on $D_-$ and $D_+$ branes is
defined by the following equal-time quantum (anti-)commutation
relations
\bea \la{commutator-open1}
&&[X^I(\sigma,\tau), \dot{X}^J(\sigma^\prime,\tau)]=
i\pi|\a| \delta^{IJ} \delta(\sigma-\sigma^\prime),\\
\la{commutator-open2}
&& \{S^{Aa}(\sigma,\tau), S^{Bb}(\sigma^\prime,\tau) \}=
\half \pi|\a| \delta^{ab} \delta^{AB} \delta(\sigma-\sigma^\prime).
\eea
For an open string, we have the boundary
conditions on the fields, Eqs. \eq{bc-boson}-\eq{bc-fermion}.
These boundary conditions will be incorporated in the mode
expansion of the fields.

\subsubsection{$D_-$-brane}

We first discuss the quantization of open strings on $D_-$-brane.
The mode expansion of the bosonic coordinates satisfying the
boundary condition and the equation of motion is given by
\bea \la{open-boson}
&& X^r(\sigma, \tau)=\cos\mu\tau x_0^r + \sin\mu\tau \frac{p_0^r}{\mu} +i\sum_{n
\neq 0}\frac{1}{\omega_n}\a_n^{r} e^{-i\omega_n \tau}
\cos\frac{n\sigma}{|\a|}, \xx
&& X^{r^\prime}(\sigma, \tau)= x_0^{r^\prime}(\sigma) +
\sum_{n \neq 0}\frac{1}{\omega_n}\a_n^{r^\prime} e^{-i\omega_n \tau}
\sin\frac{n\sigma}{|\a|},
\eea
where the zero mode part, $x_0^{r^\prime}(\sigma)$, represents
a D-brane position located either at the origin or away from the
origin and is given by
\be \la{x0}
x_0^{r^\prime}(\sigma)=\frac{x_0^{r^\prime}}{1+e^{\mu|\a|\pi}}
(e^{\mu \sigma} + e^{\mu(|\a|\pi-\sigma)}).
\ee
Similarly the mode expansion of the fermion is found to be
\bea \la{open-fermion}
&& S^1 (\sigma,\tau)= \cos \mu\tau S_0
- \sin\mu\tau \Omega\Pi S_0
+ \sum_{n \neq 0}c_n(\varphi_n^1(\sigma,\tau) \Omega S_n
+i \rho_n \varphi_n^2(\sigma,\tau)\Pi S_n), \xx
&& S^2 (\sigma,\tau)= \cos \mu\tau \Omega^T S_0
- \sin\mu\tau \Pi S_0
+ \sum_{n \neq 0}c_n(\varphi_n^2(\sigma,\tau) S_n
- i \rho_n \varphi_n^1(\sigma,\tau)\Pi \Omega S_n).
\eea
One can check that the mode expansion \eq{open-fermion} satisfies
the boundary condition \eq{bc-fermion} and \eq{bc-sigma1} as well
as the equation of motion \eq{eom-fermion}.

The commutation relations for the modes in Eqs. \eq{open-boson}
and \eq{open-fermion} are determined by the field quantization in
\eq{commutator-open1} and \eq{commutator-open2} to be respectively
\bea \la{comm-open}
&& [x_0^r, p_0^s]=i\delta^{rs}, \qquad [\a_n^{I},\a_m^{J}]= \omega_n
\delta_{m+n,0} \delta^{IJ}, \\
&& \{S_n^{a}, S_m^{b} \} = \frac{1}{4} \delta_{n+m,0}
\delta^{ab}, \qquad (n, m \in \mathbf{Z}).
\eea

\subsubsection{$D_+$-brane}

The mode expansion of the bosons is exactly the same as the
$D_-$-branes, Eq. \eq{open-boson}, and for the fermions we find
that
\bea \la{open-fermion2}
&& S^1 (\sigma,\tau)= \cosh \mu\sigma S_0
+ \sinh \mu\sigma \Omega\Pi S_0
+ \sum_{n \neq 0}c_n(\varphi_n^1(\sigma,\tau) \widetilde{S}_n
+i \rho_n \varphi_n^2(\sigma,\tau)\Pi S_n), \xx
&& S^2 (\sigma,\tau)= \cosh \mu\sigma \Omega^T S_0
+ \sinh \mu\sigma \Pi S_0
+ \sum_{n \neq 0}c_n(\varphi_n^2(\sigma,\tau) S_n
- i \rho_n \varphi_n^1(\sigma,\tau)\Pi \widetilde{S}_n),
\eea
where
\be \la{sn}
\widetilde{S}_n = \frac{1}{\omega_n}\Bigl(\frac{n}{|\a|} \Omega
- i \mu \Pi \Bigr) S_n.
\ee
One can check that the mode expansion \eq{open-fermion2} satisfies
the boundary conditions, Eqs. \eq{bc-fermion} and \eq{bc-sigma2},
as well as the equation of motion, Eq. \eq{eom-fermion}.
For the $D_+$-branes, the zero modes of the fermions depend on the
worldsheet space coordinate $\sigma$, and so there is no direct relation
with the zero modes of the closed string \ct{skenderis1}.

The anti-commutation relation of the modes for $D_+$-branes is not
trivial and some technical details are given in Appendix B:
\bea \la{d+10}
&& \{S_0^{a}, S_0^{b} \} = \frac{\pi \mu |\a|}{4\sinh \pi \mu |\a|}
\Bigl(\delta^{ab}\cosh\pi\mu|\a|-(\Omega\Pi)^{ab}\sinh\pi\mu|\a|
\Bigr),\\
\la{d+1n0}
&& \{S_n^{a}, S_m^{b} \} = \frac{1}{4} \delta_{n+m,0}
\delta^{ab}, \qquad (n,m \neq 0).
\eea
Note that the result \eq{d+10} has the correct flat space limit,
$\mu \to 0$, including coefficient.
Using the relation \eq{sn} for the nonzero mode $\widetilde{S}_n$, the following
anti-commutation relations can be derived from Eq. \eq{d+1n0}:
\bea \la{d+t1}
&& \{\widetilde{S}_n^{a}, \widetilde{S}_m^{b} \} = \frac{1}{4} \delta_{n+m,0}
\delta^{ab}, \qquad (n,m \neq 0), \\
\la{d+t2}
&& \{\widetilde{S}_n^{a}, S_m^{b} \}=
 \frac{1}{4\omega_n}\Bigl(\frac{n}{|\a|} \Omega_{ab}
-i \mu \Pi_{ab} \Bigr)\delta_{n+m,0}.
\eea

As shown in next section, for the D5-branes of type $(4,0)$ or
$(0,4)$ to be supersymmetric, a nontrivial flux of gauge field is
necessarily turned on in the worldvolume. The inclusion of the
gauge field corresponds to the addition of the following boundary
term
\be \la{bi-boundary}
S_B = \frac{1}{2\pi \a^\prime p^+} \int d\tau \int_{0}^{2\pi \a^\prime |p^+|}
d\sigma F^{-r} \p_\sigma X^r =
\frac{\mu}{4 \pi \a^\prime p^+} \int_{\p \Sigma}  d\tau X^r X^r,
\ee
where the Born-Infeld flux $F^{-r}$ is given by
\be \la{bi}
F^{-r}= \mu X^r.
\ee
This affects the Neumann boundary condition and the appropriate
boundary condition turns out to be \ct{gaberdiel}
\be \la{bc-d+}
\p_\sigma X^r|_{\p \Sigma} = \mu X^r|_{\p\Sigma}.
\ee
The boundary coupling in Eq. \eq{bi-boundary} is indeed
generated by the superpotential in the $\CN=(2,2)$
worldsheet supersymmetric theory \ct{hikida}.
The mode expansion of the Neumann coordinates $X^r(\sigma,
\tau)$ is then given by
\bea \la{mode-d+}
X^r(\sigma, \tau) &=& \sqrt{\frac{2 \pi \mu |\a|}{e^{2 \pi \mu
|\a|}-1}} (x_0^r + p_0^r \tau) e^{\mu \sigma}+ i \sum_{n \neq 0}
\frac{n}{\omega_n (n-i \mu |\a|)}
\a_n^r e^{-i \omega_n \tau} \cos\frac{n \sigma}{|\a|} \xx
&& + i \sum_{n \neq 0} \frac{\mu |\a|}{\omega_n (n-i \mu
|\a|)} \a_n^r e^{-i \omega_n \tau} \sin \frac{n \sigma}{|\a|}.
\eea
The commutation relation between the modes is given by Eq.
\eq{comm-open} as usual. See Appendix B for the derivation.

\section{Supersymmetry Algebra in Plane Wave}

We now study the basic symmetry algebra of the light-cone
superstring in the plane wave described by the action \eq{gs-action}.
The thirty bosonic symmetries are generated by the ten translation
generators, $P^- = H, \;P^+, \; P^I$, the eight boost generators,
$J^{+I}$, the six $\so$ rotation generators, $J^{ij}$,
and the six $\sop$ rotation generators, $J^{i^\prime j^\prime}$.
There are also 32 supersymmetries. In the light-cone gauge the 32
components of the supersymmetries decompose into `dynamical' and
`kinematical' components. The dynamical supercharges, $Q^{-A}_{\ad}$,
commutes with the light-cone hamiltonian, but the kinematical
supercharges, $Q^{+A}_a$, for closed strings and $D_-$-branes do
not due to a fermionic mass term. However we will show that all
supersymmetries commute with the hamiltonian for $D_+$-branes
since the fermionic mass term identically vanishes.

Using the N\"other method, the symmetry generating charges can be
obtained from conserved currents. For the superstring in the plane
wave background described by the action \eq{gs-action}, the super-N\"other
charges were obtained by Metsaev \ct{metsaev1}:
\bea \la{charge-p}
&& P^+= p^+, \qquad P^I= \frac{1}{2\pi \a^\prime p^+}
\int_{0}^{2\pi \a^\prime |p^+|} d\sigma \Bigl(
\dot{X}^I \cos \mu \tau + \mu X^I \sin \mu\tau \Bigr), \\
\la{charge-h}
&& 2p^+ H= \frac{1}{2\pi \a^\prime p^+}
\int_{0}^{2\pi \a^\prime |p^+|} d\sigma \Bigl[
\half (\dot{X}_I^2 + {X^\prime_I}^2 + \mu^2 X_I^2) + i S^A \dot{S}^A
\Bigr], \\
\la{charge-j+}
&& J^{+I}= \frac{1}{2\pi \a^\prime}
\int_{0}^{2\pi \a^\prime |p^+|} d\sigma \Bigl(
\frac{\dot{X}^I}{\mu} \sin \mu \tau - X^I \cos \mu\tau \Bigr), \\
\la{charge-j1}
&& J^{ij}= \frac{1}{2\pi \a^\prime p^+}
\int_{0}^{2\pi \a^\prime |p^+|} d\sigma \Bigl[(
X^i \dot{X}^j-X^j \dot{X}^i)-\frac{i}{2} S^A \gamma^{ij} S^A \Bigr], \\
\la{charge-j2}
&& J^{i^\prime j^\prime}= \frac{1}{2\pi \a^\prime p^+}
\int_{0}^{2\pi \a^\prime |p^+|} d\sigma \Bigl[(
X^{i^\prime} \dot{X}^{j^\prime}-X^{j^\prime} \dot{X}^{i^\prime})
- \frac{i}{2} S^A \gamma^{i^\prime j^\prime} S^A \Bigr], \\
\la{charge-q+1}
&& Q^+ = \frac{\sqrt{2p^+}}{2\pi \a^\prime p^+}
\int_{0}^{2\pi \a^\prime |p^+|} d\sigma e^{i\mu\tau \Pi}(S^1 +i S^2), \\
\la{charge-q+2}
&& \bar{Q}^+ = \frac{\sqrt{2p^+}}{2\pi \a^\prime p^+}
\int_{0}^{2\pi \a^\prime |p^+|} d\sigma e^{-i\mu\tau \Pi}(S^1 - i S^2), \\
\la{charge-q-1}
&& \sqrt{2p^+}Q^{-1}= \frac{1}{2\pi \a^\prime p^+}
\int_{0}^{2\pi \a^\prime |p^+|} d\sigma \Bigl( \p_- X^I\gamma^I S^1
-\mu X_I\gamma^I \Pi S^2 \Bigr), \\
\la{charge-q-2}
&& \sqrt{2p^+}Q^{-2}= \frac{1}{2\pi \a^\prime p^+}
\int_{0}^{2\pi \a^\prime |p^+|} d\sigma \Bigl( \p_+ X^I\gamma^I
S^2 + \mu X_I\gamma^I \Pi S^1 \Bigr),
\eea
where $X^\prime_I=\frac{\p X_I}{\p \sigma}$.

In the light-cone formalism, the generators of the basic
superalgebra can be split into the kinematical generators
\be \la{kinematic}
P^+,\;\; P^I,\;\;J^{+I},\;\;J^{ij},\;\;J^{i^\prime j^\prime},\;\;
Q^{+}_a,\;\; \bar{Q}^{+}_a,
\ee
and the dynamical generators
\be \la{dynamical}
H,\;\; Q^{-1}_{\ad},\;\; Q^{-2}_{\ad}.
\ee
Note that the kinematical supersymmetry generators have $SO(8)$
positive chirality while the dynamical supersymmetry generators
have $SO(8)$ negative chirality.
The kinematical generators $P^+,\; P^I,\;\;J^{+I},\; Q^{+}_a,\; \bar{Q}^{+}_a$
depend only on the zero modes since they are effectively linear in
fields. Now we will study the supersymmetry algebra of the
symmetry generating charges, \eq{charge-p}-\eq{charge-q-2}, for the closed
and the open strings, using the mode expansion given in the
previous section.

\subsection{Closed string}

Using the mode expansion in Eqs. \eq{sol-boson} and
\eq{sol-fermion}, we get
\bea \la{gen-closed-zero}
&& P^+ =p^+, \quad P^I =p_0^I,\quad J^{+I}= -x_0^Ip^+, \\
&& Q^+=\sqrt{2p^+}(S_0^1+iS_0^2)= \sqrt{2p^+}S_0^+,
\quad \bar{Q}^+=\sqrt{2p^+}(S_0^1-iS_0^2)= \sqrt{2p^+}S_0^-.
\eea
The rotation generators $J^{IJ}=(J^{ij} \in \so, J^{i^\prime j^\prime} \in
\sop)$ are given by
\be \la{gen-closed-j}
J^{IJ} = x_0^I p_0^J - x_0^J p_0^I - \frac{i}{2} S_0^A \gamma^{IJ}
S_0^A - i \sum_{n \neq 0} \Bigl\{ \frac{1}{\omega_n}
(\a_{-n}^{AI} \a_n^{AJ} - \a_{-n}^{AJ} \a_n^{AI})
+ \half S_{-n}^A \gamma^{IJ} S_n^A \Bigr\}.
\ee
For the dynamical generators, we have
\bea
&& 2p^+ H = \half (p_{0I}^2 + \mu^2 x_{0I}^2) +2i \mu S_0^1 \Pi S_0^2 +
\sum_{n \neq 0} \Bigl\{ \a_{-n}^{AI} \a_{n}^{AI} + \omega_n
S_{-n}^A S_n^A \Bigr \}, \\
&& \sqrt{2p^+}Q^{-1} = p_{0I} \gamma^I S_0^1 - \mu  x_{0I} \gamma^I \Pi S_0^2
+ 2 \sum_{n \neq 0} \Bigl\{ c_n \a_{-n}^{1I} \gamma^I S_{n}^1
+ \frac{i\mu}{2c_n \omega_n} \a_{-n}^{2I} \gamma^I \Pi S_n^2 \Bigr\}, \\
\la{gen-closed-dyn}
&& \sqrt{2p^+}Q^{-2} = p_{0I} \gamma^I S_0^2 + \mu  x_{0I} \gamma^I \Pi S_0^1
+ 2 \sum_{n \neq 0} \Bigl\{ c_n \a_{-n}^{2I} \gamma^I S_{n}^2
- \frac{i\mu}{2c_n \omega_n} \a_{-n}^{1I} \gamma^I \Pi S_n^1
\Bigr\}.
\eea

Now, from Eqs. \eq{gen-closed-zero}-\eq{gen-closed-dyn},
it is straightforward to derive the supersymmetry algebra of the
closed string in the plane wave background \ct{metsaev2} using the
(anti-)commutation relations, \eq{commutator-mode-boson}
and \eq{commutator-mode-fermion}. (Or one can directly calculate it
from the super-N\"other charges, Eqs.
\eq{charge-p}-\eq{charge-q-2}, using the (anti-)commutation
relations, \eq{commutator-field-boson} and
\eq{commutator-field-fermion}.) Here we will present only
the non-vanishing (anti-)commutation
relations involved with odd generators, $Q^{\pm}$ and
$\bar{Q}^{\pm}=(Q^{\pm})^\dagger$, where $Q^-_{\ad}= Q^{-1}_{\ad} + i Q^{-2}_{\ad}$
and $\bar{Q}^-_{\ad} = Q^{-1}_{\ad} - i Q^{-2}_{\ad}$: \footnote{For the
commutation relations between even generators, see
\ct{metsaev1,metsaev2}.}
\bea \la{closed-jq}
&& [J^{ij}, Q^{\pm}]=\frac{i}{2} \gamma^{ij} Q^{\pm},
\qquad [J^{i^\prime j^\prime}, Q^{\pm}]=
\frac{i}{2} \gamma^{i^\prime j^\prime}Q^{\pm}, \\
&& [J^{+I}, Q^-]= - \frac{i}{2} \gamma^I Q^+, \\
&& [P^I, Q^-]=\frac{\mu}{2 p^+} \gamma^I \Pi Q^+,
\qquad [H, Q^+]= - \frac{\mu}{2p^+} \Pi Q^+,
\eea
together with the commutators that follow from these by complex
conjugation and
\bea
&& \{Q^+_a, \bar{Q}^+_b\}=\delta_{ab} 2P^+, \\
&& \{Q^+_a, \bar{Q}^-_{\ad}\}= \gamma_{a\ad}^I P^I +i\frac{\mu}{p^+}
(\Pi\gamma^I)_{a\ad} J^{+I}, \\
\la{closed-dyn-qq}
&& \{Q^-_{\ad}, \bar{Q}^-_{\bd}\}= \delta_{\ad\bd} 2H - i\frac{\mu}{2p^+}
(\gamma^{ij} \Pi)_{\ad\bd} J^{ij} + i \frac{\mu}{2p^+}
(\gamma^{i^\prime j^\prime} \Pi)_{\ad\bd} J^{i^\prime j^\prime}.
\eea
The supersymmetry algebra \eq{closed-dyn-qq} is particularly
of importance in describing string interactions in light-cone
string field theory formalism \ct{spradlin1}.
Note that the supersymmetry algebra
in \eq{closed-dyn-qq} distinguishes the two $\so$ directions
with opposite sign.\footnote{In order to derive Eq. \eq{closed-dyn-qq},
one may use the formula, \eq{fierz1}, \eq{pi-omega1},
and \eq{gamma-rel1}. The relative sign is indeed due to Eqs.
\eq{pi-omega1} and \eq{gamma-rel1}.}
This sign flip may be read off from Eq. (2.13) in \ct{metsaev1} or Eq. (2.59)
in \ct{metsaev2} since $\Pi_{\ad \bd}=- \Pi_{\ad \bd}^\prime$ in
the space with $SO(8)$ negative chirality as explained in Appendix
A. The dynamical supersymmetry algebra in Eq. \eq{closed-dyn-qq}
is consistent with the $Z_2$ symmetry which interchanges
simultaneously the two $SO(4)$ directions \ct{chu1}
\be \la{z2}
Z_2 : (x^1,x^2,x^3,x^4) \leftrightarrow (x^5,x^6,x^7,x^8).
\ee
This should be the case since the worldsheet string action
\eq{gs-action} and the hamiltonian $H$ are $Z_2$ invariant.
Indeed the plane wave supergravity spectrums in \ct{metsaev2}
explicitly respect this $Z_2$ symmetry (see also footnote 2 in
\ct{chu1}). The presence of the relative sign in Eq. \eq{closed-dyn-qq}
is different from \ct{spradlin1,pankiewicz2}
although these parts are not corrected by the interaction, so do
not affect their results.

\subsection{Open string}

In the presence of D-brane of type $(m,n)$ in the plane wave
background \eq{pp-metric}, the symmetries of open string on the
D-branes are further broken by the boundary conditions of the open
string. First of all, the translation and the boost to the
transverse directions of D-brane, generated by $P^{r^\prime}$ and
$J^{+ r^\prime}$, are no longer the symmetries of open string
modes. Furthermore the rotational symmetry $\so \times \sop$ is
more broken to $SO(m) \times SO(4-m) \times SO(n) \times SO(4-n)$.
Thus we will get $J^{r s^\prime}=0$ as expected.\footnote{However,
the translation and the boost generated by $P^{r^\prime}$ and
$J^{+ r^\prime}$ are still the symmetries of the action \eq{gs-action}
and the isometry of the target spacetime. Thus the symmetry
transformation by the broken generators, $P^{r^\prime}$ and
$J^{+ r^\prime}$ as well as $J^{r s^\prime}$, results in new
D-branes, symmetry related D-branes, which are in general
time-dependent branes. See \ct{skenderis1,skenderis2}
for a detailed discussion on the
symmetry related D-branes.}

The supersymmetry breaking is more complicated depending on a
specific boundary condition, that is, $D_{\pm}$ boundary conditions.
It was shown in \ct{skenderis1,skenderis2} that $D_-$-branes preserve
8 kinematical and 8 dynamical supersymmetries and $D_+$-branes
preserve 8 kinematical supersymmetries regardless of location.
However, it was conjectured in \ct{gaberdiel} that, among $D_+$-branes,
D1-branes and D5-branes of type $(4,0)$ or $(0,4)$ only preserve 8
dynamical supersymmetries. Here we will give a worldsheet
derivation for this conjecture.

Since the open string action is just defined by the closed string
action imposed the open string boundary conditions
\eq{bc-boson} and \eq{bc-fermion},
the super-N\"other charges of an open string will be given by a
subset of the symmetries of the closed string action which are
compatible with the open string boundary conditions.
Due to the boundary condition \eq{bc-fermion}, it turns out that
the conserved dynamical supercharge is given by the combination
\be \la{dynamical-susy}
q^- = Q^{-2} - \Omega^T Q^{-1},
\ee
or equivalently,
\be \la{dynamical-susy-2}
q^- = Q^{-1} - \Omega Q^{-2}.
\ee
Using the equations of motion, Eqs. \eq{eom-boson} and \eq{eom-fermion},
it is not difficult to show that the dynamical supercharge density $q^-_\tau$
in Eq. \eq{dynamical-susy} satisfies the following conservation law
\be \la{conser-law}
\frac{\p q^-_\tau}{\p \tau} = - \frac{\p q^-_\sigma}
{\p \sigma},
\ee
where
\bea \la{ds--current}
q^-_\sigma &=& \sqrt{\frac{1}{2p^+}}
\Bigl( (\p_\tau X^r \gamma^r \Omega^T + \mu X^r \gamma^r \Pi)(S^1 - \Omega S^2)
-(\p_\tau X^{r^\prime} \gamma^{r^\prime} \Omega^T
- \mu X^{r^\prime} \gamma^{r^\prime} \Pi)(S^1 + \Omega S^2) \xx
&& -\p_\sigma X^r \gamma^r \Omega^T (S^1 + \Omega S^2)
+\p_\sigma X^{r^\prime} \gamma^{r^\prime} \Omega^T (S^1 - \Omega S^2) \Bigr)
\eea
for $D_-$-branes and
\bea \la{ds+-current}
q^-_\sigma &=& \sqrt{\frac{1}{2p^+}}
\Bigl( - (\p_\sigma X^r \gamma^r \Omega^T - \mu X^r \gamma^r \Pi) (S^1 + \Omega S^2)
+( \p_\sigma X^{r^\prime} \gamma^{r^\prime} \Omega^T + \mu X^{r^\prime} \gamma^{r^\prime} \Pi)
(S^1 - \Omega S^2) \xx
&& + \p_\tau X^r \gamma^r \Omega^T (S^1 - \Omega S^2)
-\p_\tau X^{r^\prime} \gamma^{r^\prime} \Omega^T
(S^1 + \Omega S^2) \Bigr)
\eea
for $D_+$-branes. The open string boundary conditions,
\eq{bc-boson} and \eq{bc-fermion}, imply that, in the case of
$D_-$-branes,
\be \la{dsc-}
q^-_\sigma|_{\p \Sigma} = \sqrt{\frac{2}{p^+}}
\mu (X^{r^\prime} \gamma^{r^\prime} \Pi S^1)|_{\p \Sigma},
\ee
while, for $D_+$-branes,
\be \la{dsc+}
q^-_\sigma|_{\p \Sigma} = \sqrt{\frac{2}{p^+}}
(\p_\sigma X^r \gamma^r \Omega^T S^1- \mu X^r \gamma^r \Pi S^1 )|_{\p
\Sigma}.
\ee
Thus, in order for the dynamical supercharge
$ q^- = \frac{1}{\pi |\a|}\int d\sigma q^-_\tau$ in Eq. \eq{dynamical-susy} or Eq.
\eq{dynamical-susy-2} to be conserved, the Dirichlet coordinates
of $D_-$-branes should satisfy
\be \la{bc-}
X^{r^\prime}|_{\p \Sigma}=0, \qquad \forall \; {r^\prime}
\ee
and the Neumann coordinates of $D_+$-branes should satisfy
\be \la{bc+}
(\p_\sigma X^r \Omega^T S^1- \mu X^r \Pi S^1 )|_{\p
\Sigma}=0, \qquad \forall \; r.
\ee

One can see from Eq. \eq{bc-} that $D_-$-branes located at a constant
transverse position $x^{r^\prime}_0$ superficially appear to break
all dynamical supersymmetries,
and the violating terms vanish when the transverse position is set
to zero, $x^{r^\prime}_0=0$. However it can be shown \ct{skenderis1}
that the broken dynamical supersymmetries can be restored by
incorporating a worldsheet symmetry realized in the action
\eq{gs-action}. We will explain along the line given in
\ct{skenderis1} how to use the worldsheet symmetry
to restore the dynamical supersymmetry.

The superstring action \eq{gs-action} in light-cone gauge is
quadratic in the fields. This peculiar property implies that it is
invariant up to boundary terms under a shift of the field by a
parameter that satisfies the free field equation. For
definiteness, let us consider an arbitrary shift of the fields
\be \la{shift}
\delta X^I(\sigma, \tau) = \eta^I (\sigma, \tau), \qquad
\delta S^A(\sigma, \tau) = \epsilon^A (\sigma, \tau).
\ee
Under the transformation, the action \eq{gs-action} changes as
follows
\bea \la{variation}
\delta S &=& - \frac{1}{2\pi \a^\prime p^+} \int d\tau \int_{0}^{2\pi \a^\prime |p^+|}
d \sigma \Bigl(X^I (\p_+\p_- \eta^I + \mu^2 \eta^I)
+ 2i \bar{S}(\rho^A \p_A - \mu \Pi) \epsilon \Bigr) \xx
&& - \frac{1}{2\pi \a^\prime p^+} \int_{\p \Sigma} d\tau
\Bigl( X^I \p_\sigma \eta^I + i S^1(\epsilon^1 - \Omega \epsilon^2)
\Bigr).
\eea
For open strings, we can get a new
symmetry corresponding to the shift of the fields, Eq. \eq{shift},
if the parameters $\eta^I$ and $\epsilon^A$ satisfy the free field equations
\bea \la{b-shift-eom}
&& \p_+\p_- \eta^I + \mu^2 \eta^I = 0, \\
\la{f-shift-eom}
&& (\rho^A \p_A - \mu \Pi) \epsilon  =0,
\eea
as well as the following boundary conditions \footnote{
For D5-brane case with the boundary term \eq{bi-boundary}, the
boundary variation for Nuemann coordinates is shifted as
$\p_\sigma \eta^r \rightarrow \p_\sigma \eta^r - \mu \eta^r$ and
the boundary condition is instead $\p_\sigma \eta^r|_{\p\Sigma}=
\mu \eta^r|_{\p\Sigma}$.}
\bea \la{bc-eta}
&& \p_\sigma \eta^r|_{\p\Sigma}=0, \\
&& \p_\tau \eta^{r^\prime}|_{\p\Sigma}=0, \\
\la{bc-epsilon}
&& \epsilon^1|_{\p\Sigma}=\Omega \epsilon^2|_{\p\Sigma}.
\eea
Note that both the equations of motion and the above boundary
conditions are identical to those satisfied by the original
fields. For closed strings, the transformation \eq{shift} is
just a trivial field redefinition.

This mechanism can be used to restore the broken dynamical
supersymmetry of $D_-$-branes located away from the origin. In
this case one can combine a closed string transformation
with a transformation of the form \eq{shift}
to obtain a good symmetry of the open string. In this way one can
find modified transformation rules by the use of worldsheet
symmetries that lead to a conserved charge \ct{skenderis1,skenderis2}.
Taking into account the extra worldsheet symmetry, the on-shell
conserved current is actually the same as the current for the
brane at origin, but with $X^{r^\prime}(\sigma, \tau)$ replaced by
$(X^{r^\prime}(\sigma, \tau)- x^{r^\prime}_0(\sigma))$ where
$x^{r^\prime}_0(\sigma)$ is given by Eq. \eq{x0}. Thus the
corresponding charge expressed in terms of oscillators is exactly
the same as that for the brane at the origin. This immediately
implies that the supersymmetry algebras are also the same.

On the other hand, one can not use the worldsheet symmetry to
restore some apparently broken dynamical supersymmetry of
$D_+$-branes since the symmetry breaking terms are now involved
with Neumann coordinates as seen in Eq. \eq{bc+}.\footnote{An
essential difference between a Dirichlet coordinate
$X^{r^\prime}(\sigma,\tau)$ and a Neumann coordinate
$X^r(\sigma,\tau)$ is that $X^{r^\prime}|_{\p \Sigma}$
at boundary depends only on zero modes $x^{r^\prime}_0$, but
$X^r|_{\p \Sigma}$ does on all modes, as proved by Eq. \eq{open-boson}.}
Only special classes of $D_+$-branes allow the condition \eq{bc+}.
These are D1-branes in which, by definition, $X^r=0$ for all $r$
and D5-branes of type $(4,0)$ or $(0,4)$ with the Born-Infeld flux
\eq{bi} where $\Omega^T S^1 = \Pi S^1$. Remaining $D_+$-branes can
not satisfy the condition \eq{bc+} and thus the dynamical
supersymmetry is not conserved.

Let us decompose a generic field $\Psi(\sigma,\tau)$ into a zero mode part
$\Psi_0(\sigma,\tau)$ and a non-zero mode part $\widetilde{\Psi}(\sigma,\tau)$.
As one can see from the explicit mode expansions in Sec. 2, the
zero mode part $\Psi_0(\sigma,\tau)$ separately satisfies the
field equations, Eqs. \eq{eom-boson}-\eq{eom-fermion},
and the boundary conditions, Eqs. \eq{bc-boson}-\eq{bc-fermion}.
Similar consideration applied to Eq. \eq{variation} thus leads to
\be \la{action-dec}
S[\Psi(\sigma,\tau)] = S[\widetilde{\Psi}(\sigma,\tau)].
\ee
Since we are interested only in the on-shell values of
super-N\"other charges, Eq. \eq{action-dec} implies that a
N\"other charge $J^G$ can also be decomposed into a zero mode part
$J^G_0$ and a non-zero mode part $\widetilde{J}^G$, that is $J^G = J^G_0 +
\widetilde{J}^G$. In particular the super N\"other charges
$\widetilde{J}^G$ can be obtained by applying
the same N\"other method to $S[\widetilde{\Psi}(\sigma,\tau)]$,
which just gives the same expressions with replacement,
$\Psi(\sigma,\tau) \rightarrow \widetilde{\Psi}(\sigma,\tau)$.
After we isolate zero mode parts in this way which possibly
contain non-periodic functions, we can apply a periodic doubling
of open strings \ct{polchinski} for the remaining non-zero mode parts
since they contain only periodic functions. The doubling trick
will be useful in the actual calculation.

\subsubsection{$D_-$-brane}

In order to derive the super-N\"other charges in terms of open
string modes, we will use a proper doubling
of the interval $[0, \pi |\a|]$ to $[0, 2\pi |\a|]$, as done in \ct{dabholkar},
such that all the classical solutions satisfy the open string
boundary conditions for the interval $[0, \pi |\a|]$ and periodic
boundary conditions for $[0, 2\pi |\a|]$.
The hamiltonian $H$ for a D-brane located away from the origin
does depend on the Dirichlet zero modes
$x^{r^\prime}_0(\sigma)$, so we first isolate the part,
denoted as $\Delta H$, for the reason explained above
and directly calculate it:
\bea \la{delta-h}
2 p^+ \Delta H &=& \frac{1}{2 \pi \a}
\int_{0}^{\pi |\a|} d\sigma (x^{\prime \, 2}_{0 r^\prime}
+ \mu^2 x_{0 r^\prime}^2), \xx
&=& \frac{\mu}{\pi \a}\frac{e^{\pi \mu |\a|}-1}{e^{\pi \mu |\a|}+1}
x_{0 r^\prime}^2.
\eea
For remaining parts, we will use the doubling trick.

It is then straightforward to get the expressions of the super-N\"other
charges in terms of the mode expansions in Eqs. \eq{open-boson} and
\eq{open-fermion}. The result is \ct{dabholkar}
\bea \la{gen-open-d-}
&& P^+ =p^+, \quad P^r =p_0^r,\quad J^{+r}= -x_0^r p^+, \\
&& Q^+=\sqrt{2p^+}(1+i\Omega^T)S_0,
\quad \bar Q^+=\sqrt{2p^+}(1-i\Omega^T)S_0, \\
&& J^{rs} = x_0^r p_0^s - x_0^s p_0^r - i S_0 \gamma^{rs} S_0
- i \sum_{n \neq 0} \Bigl\{ \frac{1}{2\omega_n}
(\a_{-n}^r \a_n^s - \a_{-n}^s \a_n^r)
+  S_{-n} \gamma^{rs} S_n \Bigr\},\\
&& J^{r^\prime s^\prime} = - i S_0 \gamma^{r^\prime s^\prime} S_0
- i \sum_{n \neq 0} \Bigl\{ \frac{1}{2 \omega_n}
(\a_{-n}^{r^\prime} \a_n^{s^\prime} - \a_{-n}^{s^\prime} \a_n^{r^\prime})
+ S_{-n} \gamma^{r^\prime s^\prime} S_n \Bigr\},\\
&& 2p^+ H= 2p^+ \Delta H + \half(p_{0r}^2 + \mu^2 x_{0r}^2) - 2\mu i S_0 \Omega \Pi S_0
+ \sum_{n \neq 0} \Bigl\{
\half \a_{-n}^I \a_n^I + 2 \omega_n S_{-n} S_n \Bigr\}, \\
&& \sqrt{2p^+}Q^{-1} = p_0^r \gamma^r S_0 + \mu x_0^r \gamma^r \Omega\Pi S_0
- \sum_{n \neq 0} \Bigl\{ c_n \a_{-n}^I \Omega \gamma^I S_{n}
- \frac{i \mu}{2c_n \omega_n}\a_{-n}^I \gamma^I \Pi S_{n} \Bigr\}, \\
&& \sqrt{2p^+}Q^{-2} = p_0^r \gamma^r \Omega^T S_0 + \mu x_0^r \gamma^r \Pi S_0
+ \sum_{n \neq 0} \Bigl\{ c_n \a_{-n}^I \gamma^I S_{n}
- \frac{i \mu}{2c_n \omega_n}\a_{-n}^I \Omega^T \gamma^I \Pi S_{n}
\Bigr\},
\eea
where $J^{rs} \in SO(m)$ or $SO(n)$ and $J^{r^\prime s^\prime} \in SO(4-m)$
or $SO(4-n)$. Note that $Q^+$ and $\bar Q^+$ as well as $Q^{-1}$ and
$Q^{-2}$ are not independent of but are related to each other
since
\be
Q^+ + \bar Q^+ + i \Omega (Q^+ - \bar Q^+) =0, \qquad
Q^{-1}+ \Omega Q^{-2} =0.
\ee
Thus we take the following independent supercharges, which are
preserved supersymmetries for $D_-$-branes as shown before,
\bea \la{open-supercharges}
&& q^+ = \half \Bigl(Q^+ + \bar Q^+ - i \Omega (Q^+ - \bar Q^+) \Bigr)
= 2 \sqrt{2 p^+} S_0, \\
&& q^- = Q^{-2} - \Omega^T Q^{-1} = 2 Q^{-2}.
\eea

Similarly we present only the non-vanishing (anti-)commutation
relations involved with the odd generators, $q^{\pm}$:
\bea \la{open-jq}
&& [J^{rs}, q^{\pm}]=\frac{i}{2} \gamma^{rs} q^{\pm},
\qquad [J^{r^\prime s^\prime}, q^{\pm}]=
\frac{i}{2} \gamma^{r^\prime s^\prime}q^{\pm}, \\
&& [J^{+r}, q^-]= \frac{i}{2} \Omega^T \gamma^r q^+, \\
&& [P^r, q^-]=-i \frac{\mu}{2 p^+} \gamma^r \Pi q^+,
\qquad [H, q^+]= i \frac{\mu}{2p^+} \Omega \Pi q^+, \\
&& \{q^+_a, q^+_b\}=\delta_{ab} 2P^+, \\
&& \{q^+_a, q^-_{\ad}\}= (\Omega \gamma^r)_{a\ad}  P^r - \frac{\mu}{p^+}
(\Pi \gamma^r)_{a\ad} J^{+r}, \\
 \la{open-dyn-qq}
&& \{q^-_{\ad}, q^-_{\bd}\}= \delta_{\ad\bd} 2(H-\Delta H) + \frac{\mu}{2p^+}
\Bigl( (\gamma^{rs}_I \Pi\Omega)_{\ad\bd} J^{rs}_I +
(\gamma^{r^\prime s^\prime}_I \Pi\Omega)_{\ad\bd} J^{r^\prime s^\prime}_I
\Bigr) \xx
&& \qquad \qquad \;\; - \frac{\mu}{2p^+}
\Bigl( (\gamma^{rs}_{II} \Pi\Omega)_{\ad\bd} J^{rs}_{II} +
(\gamma^{r^\prime s^\prime}_{II} \Pi\Omega)_{\ad\bd} J^{r^\prime s^\prime}_{II} \Bigr),
\eea
where $J^{rs}_I \in SO(m), J^{r^\prime s^\prime}_I \in SO(4-m),
J^{rs}_{II} \in SO(n)$, and $J^{r^\prime s^\prime}_{II} \in SO(4-n)$.
Note that the open string supersymmetry algebra
in \eq{open-dyn-qq} also distinguishes the two $\so$ directions
with relative sign. The supersymmetry algebra \eq{open-dyn-qq}
shows that the BPS bound is saturated by the states with energy
$\Delta H$. This implies that the brane located away from the
origin does not tend to move towards the origin \ct{skenderis2}.

\subsubsection{$D_+$-brane}

As identified by Skenderis and Taylor \ct{skenderis1,skenderis2},
the open strings on $D_+$-brane preserve a different kind of
kinematical supersymmetries not descending from the closed string.
\footnote{We thank K. Skenderis and M. Taylor for related discussions.}
The conserved N\"other current is
\bea \la{current-d+}
&& q^+_\tau = \sqrt{2p^+} \sqrt{\frac{\pi \mu |\a|}{\sinh \pi \mu |\a|}} \,
e^{\mu(\sigma-\half \pi|\a|) \Omega\Pi}(S^1 + \Omega S^2), \\
&& q^+_\sigma = \sqrt{2p^+} \sqrt{\frac{\pi \mu |\a|}{\sinh \pi \mu |\a|}} \,
e^{\mu(\sigma-\half \pi|\a|) \Omega\Pi}(S^1 - \Omega S^2).
\eea
It is simple to check that
\be \la{conser-d+}
\frac{\p q^+_\tau}{\p \tau} + \frac{\p q^+_\sigma}
{\p \sigma}=0,
\ee
using the equations of motion \eq{eom-fermion} and that
\be \la{bc-current}
q^+_\sigma|_{\p \Sigma} = 0.
\ee
This together with the current conservation implies that the
charge
\be \la{kin-charge-d+}
q^+ = \frac{1}{\pi|\a|}\int_0^{\pi|\a|} d\sigma \, q^+_\tau
\ee
is conserved and in particular all non-zero modes cancel against
each other. In Appendix C, we will directly derive the kinematical
supercurrent \eq{current-d+} from the open string mode expansion.

Using the expression for the supercharges of closed string,
Eqs. \eq{charge-q-1}-\eq{charge-q-2}, we showed that
half of the dynamical supersymmetries are preserved by D1-branes
and D5-branes of type $(4,0)$ or $(0,4)$ only.
For the supersymmetric $D_+$-branes, the $\Omega$
matrices in Eq. \eq{omega} are the following:
\bea \la{omega-d+-d1}
&& \mbox{D1-brane of type (0,0)}: \Omega_{ab}=\delta_{ab}, \quad
\Omega_{\ad\bd} = \delta_{\ad\bd}, \\
\la{omega-d+-d5+}
&& \mbox{D5-brane of type (4,0)}: \Omega_{ab} = \Pi_{ab} =
\Pi^\prime_{ab}, \quad \Omega_{\ad\bd} = \Pi_{\ad\bd} =
- \Pi^\prime_{\ad\bd}, \\
\la{omega-d+-d5-}
&& \mbox{D5-brane of type (0,4)}: \Omega_{ab} = \Pi^\prime_{ab} =
\Pi_{ab}, \quad \Omega_{\ad\bd} = \Pi^\prime_{\ad\bd} =
- \Pi_{\ad\bd}.
\eea

Since the nature of the mode expansion depends sensitively on the
particular branes on which the open string terminates, we will
discuss the supersymmetry algebra case by case. First we will
discuss D1-brane. Since the mode expansion of bosonic fields in
this case is exactly the same as that of $D_-$-branes except for
$X^r(\sigma,\tau) =0$, it is sufficient to newly calculate the fermionic parts
only. Using the methodology mentioned
in Eq. \eq{action-dec}, it is straightforward to calculate
the hamiltonian for D1-brane and the result is
\be \la{d+-hamilton}
2p^+ H= 2p^+ \Delta H + \sum_{n \neq 0}
\Bigl\{ \half \a_{-n}^I \a_n^I + 2 \omega_n S_{-n} S_n \Bigr\}, \\
\ee
Note that the open string mass term, $S_0 \Omega \Pi S_0$,
is absent, so the ground states form a degenerate supermultiplet
while the ground state of the open string for $D_-$-brane is an
unmatched boson due to the mass term. Similarly, one can obtain
the mode expansion for the rotation generators:
\be \la{d+-j1}
J^{r^\prime s^\prime} = - i \widehat{S}_0 \gamma^{r^\prime s^\prime} \widehat{S}_0
- i \sum_{n \neq 0} \Bigl\{ \frac{1}{2 \omega_n}
(\a_{-n}^{r^\prime} \a_n^{s^\prime} - \a_{-n}^{s^\prime} \a_n^{r^\prime})
+ S_{-n} \gamma^{r^\prime s^\prime} S_n \Bigr\},
\ee
where $\widehat{S}_0$ is defined by Eq. \eq{new-s0} and
their anti-commutation relation is given by Eq. \eq{anti-com-new}.

The conserved dynamical supersymmetry is given by
\eq{dynamical-susy} and, in the case of D1-brane, it is of the
form
\bea \la{d1-q-}
\sqrt{2p^+}q^{-}&=& \sqrt{2p^+}(Q^{-2}- Q^{-1}) \xx
&=& \frac{1}{\pi |\a|}
\int_{0}^{\pi |\a|} d\sigma \Bigl( \dot{X}_I \gamma^I (S^2-S^1)
+ X^\prime_I \gamma^I (S^1 + S^2)
+ \mu X_I\gamma^I \Pi (S^1 + S^2) \Bigr).
\eea
This expression coincides with \ct{skenderis1,skenderis2,gaberdiel}.
The explicit mode expansion of the supercharge \eq{d1-q-} is then
found to be
\bea \la{d1-dyn-susy}
\sqrt{2p^+}q^{-}&=& 2 \mu \sqrt{\frac{2\tanh \frac{\pi}{2}\mu |\a|}{\pi \mu
|\a|}} x_0^{r^\prime} \gamma^{r^\prime} \widehat{S}_0 \xx
&& +\sum_{n \neq 0}\Bigl( c_n
\a^{r^\prime}_{-n}\gamma^{r^\prime}(S_n + \widetilde{S}_n)-
\frac{i\mu}{2c_n \omega_n}\a^{r^\prime}_{-n}\gamma^{r^\prime}
\Pi(S_n - \widetilde{S}_n)\Bigr).
\eea
Note that the dynamical supersymmetries in this case are preserved
regardless of transverse location \ct{skenderis1}.

The supersymmetry algebra for D1-brane is closed where $P^r
=J^{+r}=0$ identically:
\bea \la{di-jq}
&& [J^{r^\prime s^\prime}, q^{\pm}]=
\frac{i}{2} \gamma^{r^\prime s^\prime}q^{\pm}, \\
&& [H, q^\pm]=0, \\
&& \{q^+_a, q^+_b\}=\delta_{ab} 2P^+, \\
&& \{ q^+_a, q^-_{\ad} \} = \sqrt{2p^+}
\sqrt{\frac{2 \mu \tanh \frac{\pi}{2}\mu |\a|}{\pi
|\a|}} x_0^{r^\prime} \gamma^{r^\prime}_{a \ad}, \\
&& \{q^-_{\ad}, q^-_{\bd}\}= \delta_{\ad\bd} 2H.
\eea

Now we will consider the D5-brane of type (0,4) for definiteness
which presumably represents the Penrose limit of a baryon vertex
(the (4,0) brane can be treated similarly). In this case the
boundary term \eq{bi-boundary} contributes to the hamiltonian
which is now given by
\be \la{d5-h}
2p^+ H= \frac{1}{\pi |\a|}
\int_{0}^{\pi |\a|} d\sigma \Bigl[
\half \dot{X}_r^2 + \half (\p_\sigma X_r - \mu X_r)^2 +
\half (\dot{X}_{r^\prime}^2 + {X^\prime_{r^\prime}}^2 + \mu^2 X_{r^\prime}^2)
+ i S^A \dot{S}^A \Bigr].
\ee
Since the mode expansion in this case is essentially the same as
the case of D1-brane except for the Neumann coordinates,
$X^r(\sigma,\tau)$ in Eq. \eq{mode-d+}, it is sufficient to newly
calculate the parts involved with the Neumann coordinates. The
result for the hamiltonian is
\be \la{d5-h-mode}
 2p^+ H= 2p^+ \Delta H + \half p_{0r}^2
+ \sum_{n \neq 0} \Bigl\{
\half \a_{-n}^I \a_n^I + 2 \omega_n S_{-n} S_n \Bigr\}.
\ee
Similarly, for the rotation generators, we get
\bea \la{d5-j-mode}
&& J^{rs} = x_0^r p_0^s - x_0^s p_0^r - i \widehat{S}_0 \gamma^{rs} \widehat{S}_0
- i \sum_{n \neq 0} \Bigl\{ \frac{1}{2\omega_n}
(\a_{-n}^r \a_n^s - \a_{-n}^s \a_n^r)
+  S_{-n} \gamma^{rs} S_n \Bigr\}, \\
&& J^{r^\prime s^\prime} = - i \widehat{S}_0 \gamma^{r^\prime s^\prime} \widehat{S}_0
- i \sum_{n \neq 0} \Bigl\{ \frac{1}{2 \omega_n}
(\a_{-n}^{r^\prime} \a_n^{s^\prime} - \a_{-n}^{s^\prime} \a_n^{r^\prime})
+ S_{-n} \gamma^{r^\prime s^\prime} S_n \Bigr\}.
\eea

For open strings on D5-brane with the Born-Infeld flux \eq{bi}, the
preserved dynamical
supersymmetry is $q^- = (Q^{-2} - \Omega^T Q^{-1})$.
For the $(0,4)$ brane, for example, the
dynamical supercharge is given by
\bea \la{d5-dyn-mode}
\sqrt{2p^+}q^{-}&=& \sqrt{2p^+}(Q^{-2} + \Pi Q^{-1}) \xx
&=& 2 p_0^r \gamma^r \Pi \widehat{S}_0 +
 2 \mu \sqrt{\frac{2\tanh \frac{\pi}{2}\mu |\a|}{\pi \mu
|\a|}} x_0^{r^\prime} \gamma^{r^\prime} \Pi \widehat{S}_0 \xx
&& + 2\sum_{n \neq 0} \Bigl\{ c_n \a_{-n}^I \gamma^I S_{n}
+ \frac{i \mu}{2c_n \omega_n}\a_{-n}^I \Pi \gamma^I \Pi S_{n}
\Bigr\}.
\eea

The supersymmetry algebra is closed as in D1-brane:
\bea \la{d5-susy-alg}
&& [J^{rs}, q^{\pm}]=
\frac{i}{2} \gamma^{rs}q^{\pm}, \qquad [J^{r^\prime s^\prime}, q^{\pm}]=
\frac{i}{2} \gamma^{r^\prime s^\prime}q^{\pm}, \\
&& [J^{+r}, q^-]= - \frac{i}{2} \gamma^r \Pi q^+, \\
&& [P^r, q^\pm]=[H, q^\pm]=0,\\
&& \{q^+_a, q^+_b\}=\delta_{ab} 2P^+, \\
&& \{ q^+_a, q^-_{\ad} \} = (\Pi \gamma^r)_{a \ad} P^r
+ \sqrt{2p^+} \sqrt{\frac{2 \mu \tanh \frac{\pi}{2}\mu |\a|}{\pi
|\a|}} x_0^{r^\prime} (\Pi \gamma^{r^\prime})_{a \ad},\\
&& \{q^-_{\ad}, q^-_{\bd}\}= \delta_{\ad\bd} 2H.
\eea
Note that the supersymmetry algebra for $q^-$ of $D_+$-branes does
not contain the angular momentum parts and all the supersymmetries
commute with the light-cone hamiltonian unlike $D_-$-branes.
Consequently the vacuum for $D_+$-branes is degenerate, containing
eight bosons and eight fermions as in flat space, and forms a
supermultiplet. The vacuum for $D_-$-branes is, however, a singlet
under the $q^-$ supersymmetry and $q^+$ is a spectrum generating
operator.

\section{Discussion}

In this paper we carefully analyzed the supersymmetry algebra of
plane wave superstrings. We found that the supersymmetry algebra
of closed string respects the $\so \times \sop \times Z_2$
symmetry where the $Z_2$ exchanges the first
$\so$ with the second $\sop$. We gave a worldsheet derivation
for conserved supersymmetries, from which we showed that half of
the dynamical supersymmetries are preserved by D1-branes and
D5-branes of type $(4,0)$ or $(0,4)$ only among $D_+$-branes. In
addition we exhaustively analyzed the supersymmetry algebra of
open strings on half BPS D-branes, both $D_-$-branes and
$D_+$-branes. We showed that the algebra is closed including a new
kind of kinematical supersymmetry restored by incorporating
worldsheet symmetry suggested by Skenderis and
Taylor \ct{skenderis1,skenderis2} and all the supersymmetries
for $D_+$-branes commute with the light-cone hamiltonian unlike
$D_-$-branes. Throughout this paper we used the 8-component
spinors, $SO(8)$ Majorana-Weyl spinors, so we hope the
supersymmetry algebras presented here will be useful for some
string field theory calculation in the plane wave background.

In this paper we considered only open strings whose end points are
attached on the same D-branes. One may consider open strings
connecting Dp-Dp$^\prime$ branes, p-p$^\prime$ strings, in the plane wave
background. The p-p$^\prime$ string was analyzed
in \ct{bergman,gaberdiel} by the boundary state description and
the computation of cylinder diagrams, mainly for instantonic
branes. It will be interesting to explicitly study the
supersymmetry algebra of p-p$^\prime$ strings as done in this paper, since
these open strings, especially being BPS states, are dual to a
defect conformal field theory in a plane wave
background \ct{dewolfe,lee2,sken-tayl}.
We will report this result elsewhere \ct{cha}.

\section*{Acknowledgments}
We thank K. Skenderis and M. Taylor for helpful correspondence
that allows us to correct an acute error in the preliminary
version of this paper. We are also grateful to Kyung-Seok Cha and
Chanyong Park for checking several parts of our calculation. BHL
thank H. Nicolai and MPI for the hospitality and Nakwoo Kim for
helpful discussions. We are supported by the grant from the
Interdisciplinary Research Program of the KOSEF (No.
R01-1999-00018) and the special grant of Sogang University in
2002. BHL was supported by the Korean Research Foundation Grant
KRF D00027.

\appendix

\section*{Appendix}

\section{Notations, Definitions and Useful Formula}

The conventions for the indices are: \\
$ I, J, K, \cdots = 1, \cdots, 8$: $SO(8)$ vector indices,\\
$ a, b, c, \cdots =1, \cdots, 8$: $SO(8)$ spinor indices with
positive chirality, \\
$ \ad, \bd, \cd, \cdots =1, \cdots, 8$: $SO(8)$ spinor indices with
negative chirality,\\
$ i, j, k, \cdots = 1, \cdots, 4$: $\so$ vector indices,\\
$ i^\prime, j^\prime, k^\prime, \cdots = 5, \cdots, 8$: $\sop$ vector indices.\\
$ A, B =1,2$: worldsheet (spinor, vector, etc.) indices.

In this paper we consider a Dp-brane of type $(m,n), \;m+n=p-1$,
with $m$ Neumann directions in $\so$ and $n$ Neumann directions in
$\sop$. Thus we distinguish the Neumann directions and the Dirichlet
directions with indices: \\
$ r, s, t, \cdots = 1, \cdots, m, 5, \cdots, 4+n$:
vector indices in Neumann directions,\\
$ r^\prime, s^\prime, t^\prime, \cdots =
m+1, \cdots, 4, 5+n, \cdots, 8$: vector indices in Dirichlet directions.

The spacetime metric is $\eta^{\mu\nu}=(-1,+1,\cdots,+1)$ where
$\mu,\nu$ are $SO(9,1)$ vector indices. We decompose $X^\mu$ into
the light-cone and transverse coordinates: $X^\mu = (X^+, X^-,
X^I)$ where
\be \la{x+-}
X^{\pm}=\frac{1}{\sqrt{2}}(X^0 \pm X^9).
\ee
The worldsheet metric is $\eta^{AB}=(-1,+1)$ where $A,B=\tau,\sigma$.

We adopt the chiral representation for $SO(9,1)$ Dirac matrices
$\Gamma^\mu$ used in \ct{metsaev1,metsaev2}
\be \la{10d-gamma}
\Gamma^\mu = \pmatrix{ 0 & \gamma^\mu \cr
                       \bar{\gamma}^\mu & 0 }
\ee
where the $16 \times 16$ matrices,
$\gamma^\mu=(\gamma^\mu)^{\a\b}=(1, \gamma^I, \gamma^9),
\;\bar{\gamma}^\mu=(\bar{\gamma}^\mu)_{\a\b}=(-1, \gamma^I,
\gamma^9), \; \a,\b=1, \cdots, 16$, satisfy
\be \la{10d-clifford}
\gamma^\mu \bar{\gamma}^\nu + \gamma^\nu \bar{\gamma}^\mu = 2
\eta^{\mu\nu} = \bar{\gamma}^\mu \gamma^\nu + \bar{\gamma}^\nu
\gamma^\mu.
\ee
The $SO(9,1)$ chirality matrix $\Gamma_{11} \equiv
\Gamma^0 \cdots \Gamma^9$ is given by
\bea \la{gamma11} &&
\Gamma_{11} = \pmatrix{ 1_{16} & 0 \cr
                       0 & -1_{16} }, \\
&& \gamma^0 \bar{\gamma}^1 \cdots \gamma^8 \bar{\gamma}^9 =
1_{16},\xx
&& \bar{\gamma}^0 \gamma^1 \cdots \bar{\gamma}^8 \gamma^9 =
-1_{16} \nonumber.
\eea

We further assume the following block decomposition for $\gamma^I$
\be \la{8d-gamma}
\gamma^I = \pmatrix{ 0 & \gamma^I_{a\ad} \cr
                       \tilde{\gamma}^I_{\ad a} & 0 }
\ee
where $8 \times 8$ matrices, $\gamma^I_{a\ad}, \; \tilde{\gamma}^I_{\ad
a}=({\gamma^I}^T)_{\ad a}$, satisfy
\bea \la{8d-clifford}
&& \gamma^I_{a\ad} \tilde{\gamma}^J_{\ad b} + \gamma^J_{a\ad} \tilde{\gamma}^I_{\ad b}
= 2\delta^{IJ}\delta_{ab}, \xx
&&\tilde{\gamma}^I_{\ad a} \gamma^J_{a \bd} + \tilde{\gamma}^J_{\ad a}
\gamma^I_{a \bd} = 2\delta^{IJ}\delta_{\ad \bd}.
\eea
Since we use $SO(8)$ chiral spinors, we take the $SO(8)$ chirality
matrix $\gamma \equiv \gamma^1\bar{\gamma}^2 \cdots \gamma^7
\bar{\gamma}^8 = \gamma^0 \bar{\gamma}^9$ as
\be \la{gamma9}
\gamma=\pmatrix{ 1_{8} & 0 \cr
                       0 & -1_{8} }.
\ee

We use the following definitions
\bea \la{pis}
&& \Pi_{ab} = \gamma^1\tilde{\gamma}^2 \gamma^3 \tilde{\gamma}^4,
\qquad \Pi_{\ad \bd}= \tilde{\gamma}^1 \gamma^2 \tilde{\gamma}^3
\gamma^4, \\
&& \Pi_{ab}^\prime = \gamma^5\tilde{\gamma}^6 \gamma^7 \tilde{\gamma}^8,
\qquad \Pi_{\ad \bd}^\prime = \tilde{\gamma}^5 \gamma^6
\tilde{\gamma}^7 \gamma^8.
\eea
Due to the normalization \eq{gamma9}, $\Pi_{ab}=\Pi_{ab}^\prime,
\; \Pi_{\ad \bd}=- \Pi_{\ad \bd}^\prime$. Note that the matrix
$\Pi$ is symmetric and traceless and $\Pi^2=1$.
We define the antisymmetrized products of gamma matrices, e.g.,
\bea \la{gammas}
&& \gamma^{IJ}_{ab}=\half (\gamma^I_{a\ad} \tilde{\gamma}^J_{\ad b}
- \gamma^J_{a\ad} \tilde{\gamma}^I_{\ad b}), \xx
&& \gamma^{IJ}_{\ad \bd}= \half ( \tilde{\gamma}^I_{\ad a} \gamma^J_{a \bd}
- \tilde{\gamma}^J_{\ad a} \gamma^I_{a \bd}), \\
&& \gamma^{IJK}_{a \bd}=\frac{1}{3!} (\gamma^I_{a\ad} \tilde{\gamma}^J_{\ad b}
\gamma^K_{b\bd} \pm \mbox{5 terms}), \xx
&& \gamma^{IJK}_{\ad b}=\frac{1}{3!} (\tilde{\gamma}^I_{\ad a} \gamma^J_{a \bd}
\tilde{\gamma}^K_{\bd b} \pm \mbox{5 terms}).
\eea
$\gamma^{IJ}$ is an antisymmetric matrix, i.e.,
$\gamma^{IJ}_{ba}=-\gamma^{IJ}_{ab}$ while
$\gamma^{IJKL}_{ba}=\gamma^{IJKL}_{ab}$. Similarly they are true
with dotted indices.

The 16-component spinor $\theta$ is decomposed in terms of the
8-component spinors as
\be \la{so8-spinor}
\theta^\a =\left(%
\begin{array}{c}
  S^a \\
  Q^{\ad} \\
\end{array}%
\right),
\ee
where $S^a=-\half \gamma^+ \bar{\gamma}^- \theta$ is a positive
chirality spinor and $Q^{\ad}=-\half \gamma^- \bar{\gamma}^+ \theta$ is a negative
chirality spinor. Here $\gamma^\pm$ is defined by
\be \la{lc-gamma}
\gamma^\pm = \frac{1}{\sqrt{2}}(\gamma^0 \pm \gamma^9).
\ee
In computing the supersymmetry algebra we need to use the Fierz
identity. For spinors $S_1$ and $S_2$ with positive chirality,
\be \la{fierz1}
S_1^a S_2^b = \frac{1}{8}\delta_{ab}S_1S_2 +
\frac{1}{16}S^1\gamma^{IJ}S_2 \gamma^{IJ}_{ab} +\frac{1}{384}S_1
\gamma^{IJKL} S_2 \gamma^{IJKL}_{ab}.
\ee
Also the similar expression is true for negative chirality
spinors. Another useful identity \ct{green2} is
\bea \la{fierz2}
\gamma^I_{a \ad} \gamma^J_{b \bd} &=& \frac{1}{8}(\delta^{IJ}
\delta_{ab}\delta_{\ad\bd} + \delta_{ab}\gamma^{IJ}_{\ad\bd}
+\delta_{\ad\bd} \gamma^{IJ}_{ab}) \xx
&+& \frac{1}{16}\delta^{IJ}(\gamma^{KL}_{ab}\gamma^{KL}_{\ad\bd}
+\frac{1}{24}\gamma^{KLMN}_{ab} \gamma^{KLMN}_{\ad\bd}) \xx
&-& \frac{1}{8}(\gamma^{IK}_{ab}\gamma^{JK}_{\ad\bd}
+\gamma^{JK}_{ab} \gamma^{IK}_{\ad\bd}) \xx
&-& \frac{1}{48}(\gamma^{IKLM}_{ab}\gamma^{JKLM}_{\ad\bd}
+\gamma^{JKLM}_{ab} \gamma^{IKLM}_{\ad\bd}) \xx
&+& \frac{1}{16}(\gamma^{IJKL}_{ab}\gamma^{KL}_{\ad\bd}
+\gamma^{KL}_{ab} \gamma^{IJKL}_{\ad\bd}).
\eea

In order to discuss the boundary condition of open strings living
on a Dp-brane, it is needed to introduce the matrix $\Omega$
defined by
\be \la{omega}
\Omega_{ab}=\biggl(\prod_{I\in \CN} \gamma^I \biggr)_{ab},
\qquad \Omega_{\ad\bd}=\biggl(\prod_{I\in \CN} \gamma^I \biggr)_{\ad\bd},
\ee
where $\CN$ denotes the set of Neumann directions. Note that
$\Omega^T \Omega =1$. Thus, for the $D_-$-branes, $\Pi\Omega$ is
an antisymmetric matrix, i.e., $(\Pi\Omega)^T = - \Pi\Omega$, while,
for $D_+$-branes, $(\Pi\Omega)^T = \Pi\Omega$. The following
commutation relations are useful in the calculation
\bea \la{pi-omega1}
&& \{ \gamma^i, \Pi \}=0, \qquad [\gamma^{i^\prime}, \Pi ]=0, \\
\la{pi-omega2}
&&  \{ \gamma^r, \Omega \}=0, \qquad [\gamma^{r^\prime}, \Omega
]=0.
\eea
We often use the subscripts $I \in \so$ and $II \in \sop$ to distinguish the two
different $\so$ directions of Neumann and Dirichlet coordinates,
e.g., $\gamma^r_I,\; \gamma^r_{II}, \; \gamma^{r^\prime}_I$, etc.
Using \eq{pi-omega1} and \eq{pi-omega2}, the
following formula can be derived
\bea \la{gamma-rel1}
&& f_{IJ} \gamma^K \gamma^{IJ} \Pi \gamma^K = f_{IJ} \gamma^K \Pi \gamma^{IJ}
\gamma^K = 4(f_{ij}\gamma^{ij}
- f_{i^\prime j^\prime}\gamma^{i^\prime j^\prime})\Pi, \\
\la{gamma-rel2}
&&  \gamma^t \gamma^{rs}_I \Pi \gamma^t = (4-m+n) \gamma^{rs}_I
\Pi, \quad \gamma^{t^\prime} \gamma^{r^\prime s^\prime}_I \Pi \gamma^{t^\prime}
= (4+m-n) \gamma^{r^\prime s^\prime}_I \Pi, \quad \mbox{etc.}
\eea
where $f_{IJ}$ is a fermion bilinear.

Finally we list useful integral formula which are used in the
calculation of $D_+$-branes:
\bea \la{integral-1}
&& \int_0^{\pi |\a|} \cosh \mu \sigma \cos \frac{n \sigma}{|\a|}\, d \sigma
= (-)^n \frac{\mu \alpha^2}{\mu^2 \alpha^2 + n^2}
\sinh \pi \mu |\a|, \\
&& \int_0^{\pi |\a|} \sinh \mu \sigma \sin \frac{n \sigma}{|\a|}\, d \sigma
= -(-)^n \frac{n|\alpha|}{\mu^2 \alpha^2 + n^2}\sinh \pi \mu
|\a|, \\
&& \int_0^{\pi |\a|} \cosh \mu \sigma \sin \frac{n \sigma}{|\a|}\, d \sigma
= - \frac{n |\alpha|}{\mu^2 \alpha^2 + n^2}
\Bigl((-)^n \cosh \pi \mu |\a| -1 \Bigr), \\
\la{integral-4}
&& \int_0^{\pi |\a|} \sinh \mu \sigma \cos \frac{n \sigma}{|\a|}\, d \sigma
= \frac{\mu \alpha^2}{\mu^2 \alpha^2 + n^2}
\Bigl((-)^n \cosh \pi \mu |\a| -1 \Bigr).
\eea

\section{(Anti-)Commutation Relations for $D_+$-branes}

To determine the anti-commutation relation of the zero modes, Eq.
\eq{d+10}, let us calculate the anti-commutator for the spinor
field $S^A(\sigma, \tau)$, Eq. \eq{commutator-open2}, using the mode
expansion \eq{open-fermion2} and the anti-commutation relations,
Eqs. \eq{d+1n0}-\eq{d+t2}, for the nonzero modes:
\bea \la{commut-st}
 \{S^{1a}(\sigma,\tau), S^{1b}(\sigma^\prime, \tau)\}
 &=& \cosh\mu(\sigma+\sigma^\prime) \{S_0^{a}, S_0^{b}\}
  + \sinh\mu(\sigma+\sigma^\prime)(\Omega\Pi)_{ac} \{S_0^{c}, S_0^{b}\}
  \xx
&&+\frac{1}{4} \delta_{ab} \sum_{n \in \mathbf{Z}}
e^{i \frac{n}{|\a|}(\sigma-\sigma^\prime)}
-\frac{1}{4} \delta_{ab} \sum_{n \in \mathbf{Z}} \frac{\mu^2}{\omega_n^2}
e^{i \frac{n}{|\a|}(\sigma+\sigma^\prime)} \xx
&& -\frac{i}{4}(\Omega\Pi)_{ab} \sum_{n \in \mathbf{Z}} \frac{\mu n /|\a|}{\omega_n^2}
e^{i \frac{n}{|\a|}(\sigma+\sigma^\prime)},
\eea
where we assumed the form of $\{S_0^{a}, S_0^{b}\}$ as $A \delta_{ab} + B
(\Omega\Pi)_{ab}$. Now we will use the following integral representation to evaluate
the infinite sums in \eq{commut-st}:
\bea \la {sum1}
\sum_{n \in \mathbf{Z}} \frac{\mu^2}{\omega_n^2}
e^{i \frac{n}{|\a|}(\sigma+\sigma^\prime)} &=& - \int_C
\frac{dz}{1-e^{2\pi i z}} \frac{\mu^2 \a^2}{z^2+\mu^2\a^2}
e^{i(\sigma+\sigma^\prime)z/|\a|} \xx
&=& \pi \mu |\a| \Bigl(\frac{e^{-\mu(\sigma+\sigma^\prime)}}
{1-e^{-2\pi \mu |\a|}}- \frac{e^{\mu(\sigma+\sigma^\prime)}}
{1-e^{2\pi \mu |\a|}} \Bigr),
\eea

\bea \la {sum2}
\sum_{n \in \mathbf{Z}} \frac{\mu n /|\a|}{\omega_n^2}
e^{i \frac{n}{|\a|}(\sigma+\sigma^\prime)} &=& - \int_C \frac{dz}{1-e^{2\pi
i z}} \frac{\mu|\a|z}{z^2+\mu^2\a^2} e^{i(\sigma+\sigma^\prime)z/|\a|} \xx
&=& i \pi \mu |\a| \Bigl(\frac{e^{-\mu(\sigma+\sigma^\prime)}}
{1-e^{-2\pi \mu |\a|}} + \frac{e^{\mu(\sigma+\sigma^\prime)}}
{1-e^{2\pi \mu |\a|}} \Bigr),
\eea
where the contour $C$ consists of two lines passing
infinitesimally above and below the real axis. One can see that
the contribution from the zero modes is exactly cancelled by the
second and the third sums in \eq{commut-st} provided that the anti-commutator
of the zero modes is given as in \eq{d+10}. Thus we have the anti-commutation relation
Eq. \eq{commutator-open2}. Similarly, one can check that
$\{S^{1a}(\sigma,\tau), S^{2b}(\sigma^\prime, \tau)\} =0 $.

Using the commutation relations in Eq. \eq{comm-open} and the
identities \eq{sum1} and \eq{sum2} together with
\bea \la {sum3}
\sum_{n \in \mathbf{Z}} \frac{n^2/\a^2}{\omega_n^2}
e^{i \frac{n}{|\a|}(\sigma+\sigma^\prime)} &=& - \int_C
\frac{dz}{1-e^{2\pi i z}} \frac{z^2}{z^2+\mu^2\a^2}
e^{i(\sigma+\sigma^\prime)z/|\a|} \xx
&=& - \pi \mu |\a| \Bigl(\frac{e^{-\mu(\sigma+\sigma^\prime)}}
{1-e^{-2\pi \mu |\a|}}- \frac{e^{\mu(\sigma+\sigma^\prime)}}
{1-e^{2\pi \mu |\a|}} \Bigr),
\eea
 one can also check that
$X^r(\sigma,\tau)$ in Eq. \eq{mode-d+} satisfies the quantization
rule \eq{commutator-open1}.

\section{Kinematical Supersymmetry for $D_+$-branes}

Here we directly derive the kinematical
supercurrent \eq{current-d+} for $D_+$-branes from the open string mode expansion.
In section 2, we showed that the
fermionic mode expansions \eq{open-fermion2} satisfy the boundary
condition \eq{bc-fermion} as well as the equations
of motion \eq{eom-fermion} and the anti-commutation
relations between the modes are given by Eqs. \eq{d+10} and
\eq{d+1n0}. Thus the N\"other charge for the kinematical
supersymmetry should be represented by some combination of $S^1$
and $S^2$ fermions in Eq. \eq{open-fermion2}. We now examine
whether there can be any combination from $S^1$ and $S^2$ fermions
which reduces to a kinematical supersymmetry generator $q^+$
satisfying the standard superalgebra
\be \la{kine-susy-d+}
\{q^+_a, q^+_b\}=\delta_{ab} 2P^+.
\ee

To proceed our argument, first note that the anti-commutation
relation in Eq. \eq{d+10} can be rewritten as
\be \la{anti-com-s0}
 \{S_0^{a}, S_0^{b} \} = \frac{\pi \mu |\a|}{4\sinh \pi \mu |\a|}
\Bigl(e^{-\pi \mu |\a| \Omega\Pi} \Bigr)_{ab}
\ee
and thus a zero mode fermion defined by
\be \la{new-s0}
\widehat{S}_0 = \sqrt{\frac{\sinh \pi \mu |\a|}{\pi \mu |\a|}}
e^{\half \pi \mu |\a| \Omega\Pi} S_0
\ee
satisfies the standard anti-commutation relation
\be \la{anti-com-new}
\{\widehat{S}_0^{a}, \widehat{S}_0^{b} \} = \frac{1}{4}
\delta_{ab}.
\ee
Thus if the supersymmetry generator $q^+$ would be given by
\be \la{dyn-q0}
q^+ = 2 \sqrt{2p^+} \widehat{S}_0,
\ee
$q^+$ in \eq{dyn-q0} then satisfies the superalgebra
\eq{kine-susy-d+}. Interestingly it turns out that such a supersymmetry
generator can be constructed from the fermions
in Eq. \eq{open-fermion2}.

Notice that the zero mode parts, $S^1_0(\sigma)$ and $S^2_0(\sigma)$ of $S^1(\sigma,
\tau)$ and $S^2(\sigma, \tau)$, respectively, are given by
\bea \la{s1-s2}
&& S^1_0(\sigma)=\sqrt{\frac{\pi \mu |\a|}{\sinh \pi \mu |\a|}} \,
e^{\mu(\sigma-\half \pi|\a|) \Omega\Pi}\widehat{S}_0, \xx
&& S^2_0(\sigma)=\sqrt{\frac{\pi \mu |\a|}{\sinh \pi \mu |\a|}} \,
\Omega^T e^{\mu(\sigma-\half \pi|\a|) \Omega\Pi}\widehat{S}_0.
\eea
Since the zero modes $S^1_0(\sigma)$ and $S^2_0(\sigma)$
satisfy $S^1_0(\sigma)= \Omega S^2_0(\sigma)$,
as obviously seen in \eq{s1-s2}, we can deduce the form
of the charge density $q^+_\tau$
for the kinematical supersymmetry whose zero mode part would be
given by \eq{dyn-q0}:
\be \la{density-q+}
q^+_\tau = A e^{k \mu(\sigma-\half \pi|\a|) \Omega\Pi}
(S^1 + \Omega S^2)(\sigma, \tau),
\ee
where $k \in {\bf Z}$.
Note that only $k=0$ case is descending from the closed string.
Due to the structure of the integrals, Eqs.
\eq{integral-1}-\eq{integral-4}, the most plausible choices are
$k=-1, 0, 1$, or explicitly,
\bea \la{candidate1}
&& q^+_\tau = \sqrt{2p^+} \sqrt{\frac{\sinh \pi \mu |\a|}{\pi \mu |\a|}} \,
e^{-\mu(\sigma-\half \pi|\a|) \Omega\Pi}(S^1 + \Omega S^2), \\
\la{candidate2}
&& q^+_\tau = \sqrt{\frac{\pi \mu |\a| p^+}{\tanh \half \pi \mu |\a|}} \,
(S^1 + \Omega S^2), \\
\la{k=1}
&& q^+_\tau = \sqrt{2p^+} \sqrt{\frac{\pi \mu |\a|}{\sinh \pi \mu |\a|}} \,
e^{\mu(\sigma-\half \pi|\a|) \Omega\Pi}(S^1 + \Omega S^2).
\eea
The first two candidates, however, obtain contributions from the
non-zero modes and so none of them is conserved. For example, the
first choice \eq{candidate1} gives us
\bea \la{second}
q^+ &=& 2 \sqrt{2p^+} \widehat{S}_0
+ \frac{4 \sqrt{2p^+}}{\pi}
\sqrt{\frac{\sinh \pi\mu |\a|}{\pi \mu |\a|}} \times \xx
&& \sum_{n \neq 0} \frac{c_n}{\omega_n}
\frac{\mu n}{\mu^2 \a^2 + n^2}\Bigl(
e^{\half \pi \mu |\a| \Omega\Pi}- (-)^n
e^{-\half \pi \mu |\a| \Omega\Pi} \Bigr)
\Bigl(\Pi S_n - i \rho_n \Omega S_n \Bigr) e^{-i\omega_n \tau}.
\eea
Here we used the integral formula \eq{integral-1}-\eq{integral-4}.
But the third choice \eq{k=1} precisely gives the conserved
kinematical supercharge
\be \la{superkin-d+}
q^+ = 2 \sqrt{2p^+} \widehat{S}_0,
\ee
first identified by Skenderis and Taylor
\ct{skenderis1,skenderis2}.

\newpage


\nc{\np}[3]{Nucl. Phys. {\bf B#1}, #2 (#3)}

\nc{\pl}[3]{Phys. Lett. {\bf B#1}, #2 (#3)}

\nc{\prl}[3]{Phys. Rev. Lett. {\bf #1}, #2 (#3)}

\nc{\prd}[3]{Phys. Rev. {\bf D#1}, #2 (#3)}

\nc{\ap}[3]{Ann. Phys. {\bf #1}, #2 (#3)}

\nc{\prep}[3]{Phys. Rep. {\bf #1}, #2 (#3)}

\nc{\ptp}[3]{Prog. Theor. Phys. {\bf #1}, #2 (#3)}

\nc{\rmp}[3]{Rev. Mod. Phys. {\bf #1}, #2 (#3)}

\nc{\cmp}[3]{Comm. Math. Phys. {\bf #1}, #2 (#3)}

\nc{\mpl}[3]{Mod. Phys. Lett. {\bf #1}, #2 (#3)}

\nc{\cqg}[3]{Class. Quant. Grav. {\bf #1}, #2 (#3)}

\nc{\jhep}[3]{J. High Energy Phys. {\bf #1}, #2 (#3)}

\nc{\hep}[1]{{\tt hep-th/{#1}}}


\end{document}